\documentclass[twocolumn]{aastex62}

\usepackage{multirow}

\graphicspath{{./}{figures/}}

\received{September 7, 2018}
\revised{September 7, 2018}
\accepted{\today}

\submitjournal{ApJ}

\shorttitle{Evidence for the First Extragalactic RL Maser in NGC 253}
\shortauthors{B\'aez-Rubio et al.}

\begin{document}

\title{Evidence for the First Extragalactic Hydrogen Recombination Line Maser in NGC 253}

\correspondingauthor{Alejandro B\'aez-Rubio}
\email{abaez@cab.inta-csic.es}

\author[0000-0002-0786-7307]{Alejandro B\'aez-Rubio}
\affil{Centro de Astrobiolog\'ia (CSIC-INTA),
              Instituto Nacional de T\'ecnica Aeroespacial
Ctra de Torrej\'on a Ajalvir, km 4
28850 Torrej\'on de Ardoz, Madrid Spain}

\author{Jes\'us Mart\'in-Pintado}
\affiliation{Centro de Astrobiolog\'ia (CSIC-INTA),
              Instituto Nacional de T\'ecnica Aeroespacial
Ctra de Torrej\'on a Ajalvir, km 4
28850 Torrej\'on de Ardoz, Madrid Spain}

\author{Fernando Rico-Villas}
\affiliation{Centro de Astrobiolog\'ia (CSIC-INTA),
              Instituto Nacional de T\'ecnica Aeroespacial
Ctra de Torrej\'on a Ajalvir, km 4
28850 Torrej\'on de Ardoz, Madrid Spain}

\author{Izaskun Jim\'enez-Serra}
\affiliation{Centro de Astrobiolog\'ia (CSIC-INTA),
              Instituto Nacional de T\'ecnica Aeroespacial
Ctra de Torrej\'on a Ajalvir, km 4
28850 Torrej\'on de Ardoz, Madrid Spain}

\begin{abstract}
We present the first detection of extragalactic hydrogen recombination line maser emission in the H26$\alpha$ transition toward the inner \mbox{13.5 pc} nuclear region of the starburst galaxy \mbox{NGC 253} using ALMA data. In regions with complex continuum emission (dust, free-free and synchrotron) we propose to use the recombination line spectral index, $\alpha_\mathrm{L}$ ($S_\mathrm{L}\cdot \Delta v \propto \nu^{\alpha_\mathrm{L}}$), between the H30$\alpha$ and the H26$\alpha$ lines to study the structure of ultra-compact \mbox{H {\sc ii}}  regions and to identify maser emission ($\alpha_\mathrm{L}> 2.1$) from ionized winds. The measured values of $\alpha_\mathrm{L}$ ranged from 1.0 to 2.9. The largest $\alpha_\mathrm{L}$ can only be explained by maser emission. The measured flux density in the H26$\alpha$ maser in \mbox{NGC 253} suggests that we are observing hundreds of stars like \mbox{MWC349A,} a prototypical stellar wind where maser emission arises from its circumstellar disk. We briefly discuss the implication of the detection of maser emission in starburst galaxies like \mbox{NGC 253.}
\end{abstract}


\section{Introduction}
High-gain hydrogen recombination line (RL) masers were first detected in 1989 at mm wavelengths toward the stellar wind source \mbox{MWC349A} \citep{MartinPintado1989}. Soon after, submm maser emission in the H26$\alpha$ lines \citep{Thum1994a} and laser emission in far-IR \citep{Strelnitski1996c,Smith1997,Thum1998} were reported. Subsequent modeling \citep{Strelnitski1996a,Strelnitski1996b} clearly hinted that the physical conditions required for inverting the populations to give rise to strong maser emission are associated to dense ionized gas like that found in ionized stellar winds with a density structure of {$\sim r^{-2}$ and continuum spectral index of $\sim0.6$. These winds are not found in most of the ultra and hypercompact \mbox{H {\sc ii}} regions, suggesting that they occur during a short stage in an evolution of young massive stars \citep{Jaffe1999}. In fact, RL masers have been actively sought in both mm and submm surveys, but they were detected only in a few galactic sources \citep{Cox1995,JimenezSerra2011,JimenezSerra2013,SanchezContreras2017,Aleman2018}. 

The strongest RL maser source ever detected, \mbox{MWC349A,} shows the prototypical stellar wind spectrum from cm to far-IR wavelengths \citep{Olnon1975,Tafoya2004} and a circumstellar disk \citep{Planesas1992,Danchi2001,MartinPintado2011,BaezRubio2013,BaezRubio2014,Zhang2017} likely reminiscent of the latest phases of massive star formation. Based on the number of RL masers found in our galaxy, one would expect that galaxies undergoing strong bursts of star formation would have, at least, some objects in the evolutionary stage with the appropriate physical conditions to show maser emission. 

\mbox{NGC 253} is a large isolated spiral galaxy, of type SAB(s)c, with its disk oriented nearly edge-on \citep{Puche1991}. Located at a distance of  \mbox{$3.5\pm0.2$ Mpc} \citep{Mouhcine2005,Rekola2005}, it is one of the closest starburst galaxies making it a prime candidate to study the starburst phenomena. \mbox{NGC 253} has been studied across the whole electromagnetic spectrum revealing the complexity of its nuclear region with multiple sources showing different nature (both thermal and non thermal). Of special relevance to understand star formation in this galaxy are the sources of thermal radio-continuum emission tracing massive star formation regions, which are expected to emit radio RL emission. Previous 1-2" resolution ALMA studies of RLs at \mbox{3 mm,} in particular of the H40$\alpha$,  suggested that the emission from the \mbox{H {\sc ii}} regions was basically optically thin with temperatures of 3700--\mbox{4500 K,} providing star formation rates consistent with the IR emission \citep{Bendo2015}. 

In this Letter we have used  ALMA data to discover the first RL maser emission in the H26$\alpha$ RL toward an extragalactic object, the nucleus of \mbox{NGC 253.} We have analyzed in detail the RL emission in the most inner \mbox{13.5 pc} region, which is thought to harbour the central AGN \citep{Ulvestad1997}. The high spatial resolution and high-sensitivity ALMA observations of the H30$\alpha$ and H26$\alpha$ have been crucial to reveal the maser emission toward a super stellar cluster (SSC). 

\section{Data reduction and results}
\label{observation_section}

We have used ALMA archival data with angular resolutions higher than \mbox{$\sim 0.35$ arcsec} and sensitivities of \mbox{0.46--0.60 mJy beam$^{-1}$} (at 231.90 and \mbox{353.62 GHz} respectively) 
at a velocity resolution of \mbox{30 km s$^{-1}$.} The high angular resolution allows us to study the most compact \mbox{H {\sc ii}} regions, resolving out much of the diffuse thermal and non-thermal emission, while the high sensitivity helps in the detection of the weak H30$\alpha$ and H26$\alpha$ RLs (at 231.90 and \mbox{353.62 GHz} respectively) toward the continuum sources reported by \cite{Leroy2018}.

Data calibration and imaging were carried out in the standard manner with the CASA package (Common Astronomy Software Applications; \citealt{McMullin2007}). The pipeline reduction scripts were modified to image the whole spectral ranges. In order to subtract the local continuum (with zero or first order polynomials), we have identified the free line channels of our regions with the help of the Statcont Method \citep{SanchezMonge2018}. The continuum subtraction and spectral analysis of the H26$\alpha$ and H30$\alpha$ RL cubes were done with the MADCUBA software (see, e.g. \citealt{MartinDomenech2017}). The continuum emission map of the inner \mbox{13.5 pc,} shown in \mbox{Fig.~\ref{figure_with_maps}} in thin dark contours, clearly shows sources 10, 11, and 13 \citep{Leroy2018}. Superimposed on the continuum map, the thick red contours in \mbox{Fig.~\ref{figure_with_maps}} show the comparison of the H30$\alpha$ and H26$\alpha$ emission in a \mbox{30 km s$^{-1}$} channel at the radial velocity of \mbox{275 km s$^{-1}$.}  There is a clear ionized source (previously unknown) in the northeast from source 10, hereafter called source 10NE, which presents very weak H30$\alpha$ emission and is not associated to any continuum peak.  \mbox{Fig.~\ref{RL_profile_H30alpha}} shows the spectral line profiles measured toward all sources. We also show the Gaussian fits in this figure (red lines), whose parameters are listed in \mbox{Table 1}. We note that the measured $v_\mathrm{LSR}$ are similar to those measured in molecular lines but their line widths are broader (about \mbox{$\Delta v\sim$65--180 km s$^{-1}$} versus \mbox{$\Delta v\leq$25 km s$^{-1}$} for the molecular lines). The profiles toward 10NE shown in \mbox{Fig.~\ref{RL_profile_H30alpha}} have been smoothed to \mbox{30 km s$^{-1}$} to improve the SNR to detect the H30$\alpha$ RL. Note that there is a slight shift in the measured $v_\mathrm{LSR}$ of the H26$\alpha$ with respect to the H30$\alpha$ RL.

\begin{figure*}
\centering
\includegraphics[width=0.93\textwidth]{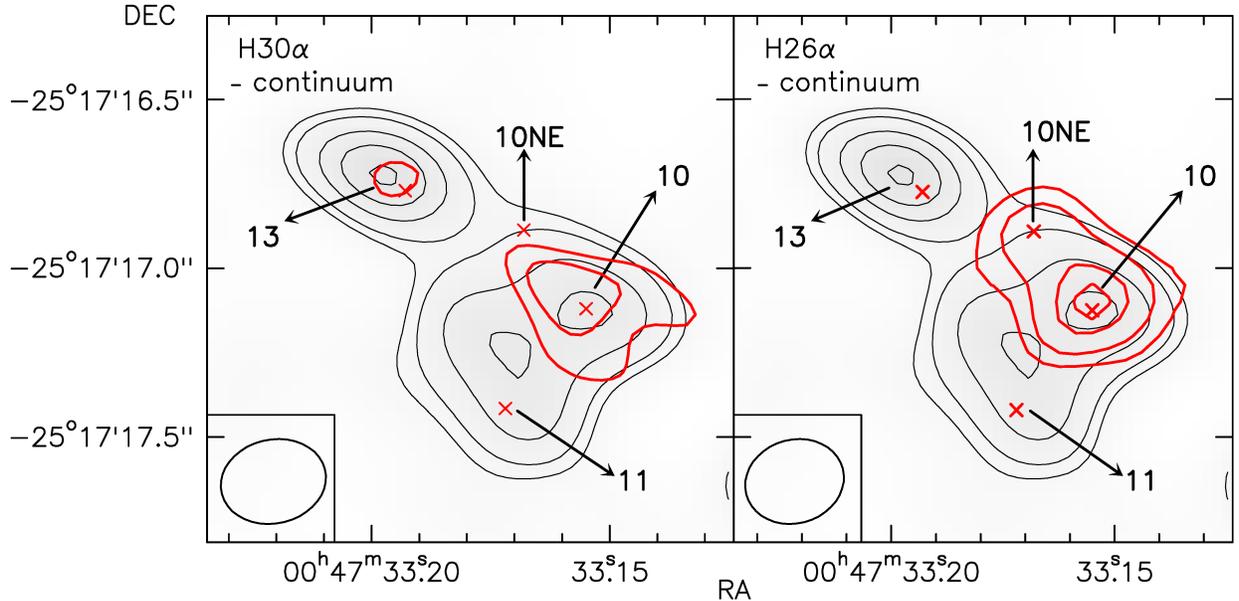}
\caption{Superimposed 231 GHz continuum (black contour levels and grayscale) and RL emission maps (red contour levels) of the H30$\alpha$ and H26$\alpha$ (left and right panels respectively) toward the inner region of the nucleus of NGC 253. The RL images were integrated in the velocity range between 260 and \mbox{290 km s$^{-1}$.} Continuum contours are -2, 2, 3, 5, and 7 times the rms noise of the image \mbox{(1 mJy)} and RL contours are -3, 3.5, 5, 7.5, and 9 times the rms in the cubes \mbox{(0.45 and 0.60 mJy for the H30$\alpha$ and H26$\alpha$ cubes respectively)}. Crosses show the RL emission sources detected in this Letter. The final beams of our observations are shown within the inset in the lower left corner.}
\label{figure_with_maps}
\end{figure*}

\begin{figure*}
\centering
\includegraphics[width=0.8\textwidth]{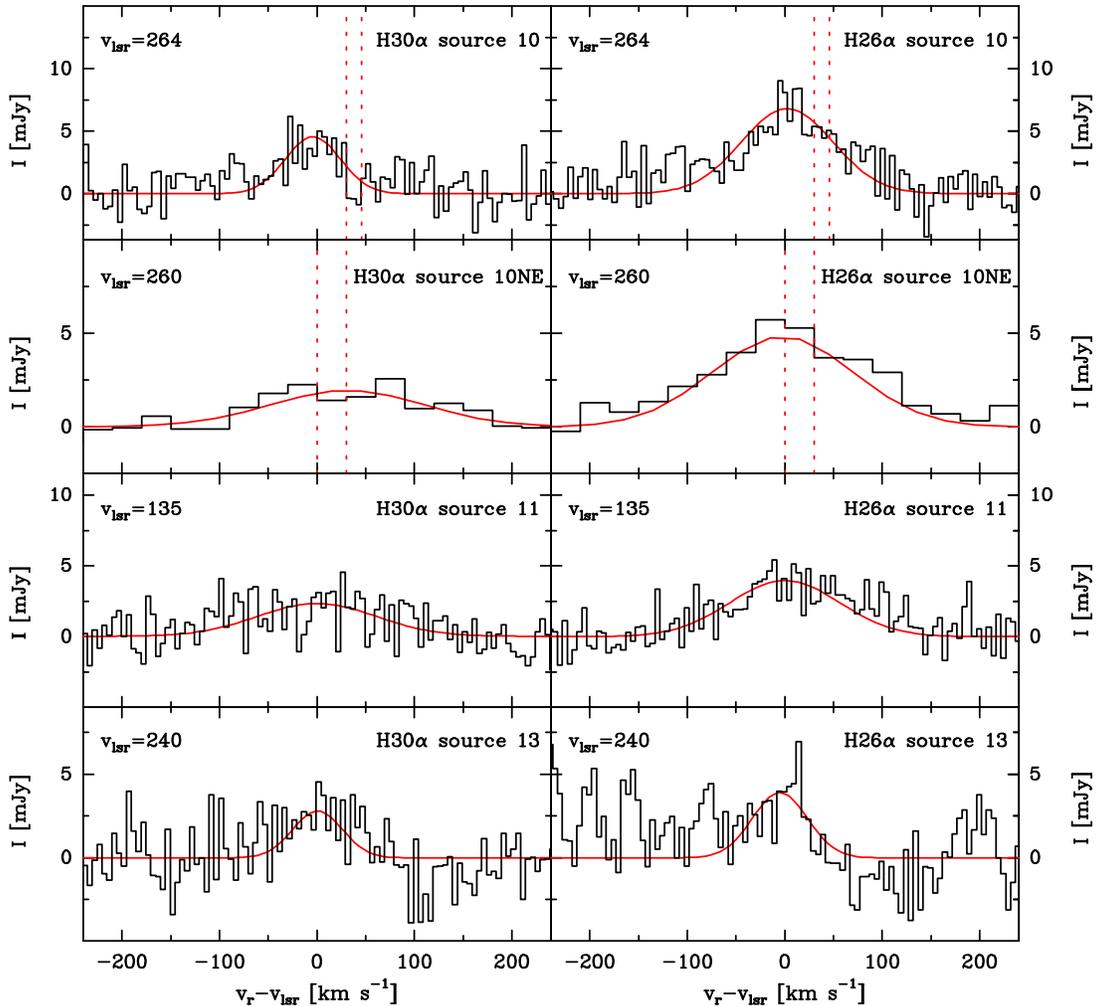}
\caption{RL profiles for H30$\alpha$ and H26$\alpha$ (left and right panels respectively) toward the compact components of NGC 253. We also show the Gaussian fits to these lines (solid red lines). The $v_\mathrm{LSR}$ of the sources are shown in the left upper part of each panel. Vertical dashed lines show the velocity ranges where there are significant differences between the intensities of the two RLs.}
\label{RL_profile_H30alpha}%
\end{figure*}

\begin{deluxetable*}{ccCrlcccc}[b!]
\tablecaption{List of compact RL sources and measured parameters: positions, detected RLs, parameters derived from Gaussian fits (velocities, peak intensities, line widths, and velocity-integrated intensities), and \mbox{RLSI}s, $\alpha_\mathrm{L}$.}
\label{table_of_spectral_indices}
\tablecolumns{9}
\tablenum{1}
\tablewidth{0pt}
\tablehead{
\colhead{Source} & \multicolumn{2}{l}{Position RL peak} 
& \colhead{Line} &  \colhead{v}&
\colhead{$S_\mathrm{l,peak}$ } 
& \colhead{$\Delta v$ } & \colhead{$\int S_\mathrm{l,peak} dv$ \tablenotemark{b}}
& \colhead{$\alpha_{\mathrm{L}}$\tablenotemark{c}}
\\
& \colhead{$\alpha$(J2000)}  &  \colhead{$\delta$(J2000)} & & \colhead{[km\ s$^{-1}$]} & \colhead{[mJy]} & \colhead{[km\ s$^{-1}$]}
& \colhead{[mJy km\ s$^{-1}$]}
}
\startdata
\multirow{2}{*}{10}& \multirow{2}{*}{00:47:33.155} & \multirow{2}{*}{-25:17:17.12}& H30$\alpha$ 
& $260.0\pm3.8$ 
& $4.55\pm0.62$ 
& $66.9\pm8.9$
& $324\pm65$ 
& \multirow{2}{*}{$2.15\pm0.69$}  \smallskip \\
 &  & & H26$\alpha$&  $266.3\pm3.4$ & $6.79\pm0.53$ & $111.0\pm8.0$ & $802\pm90$  \medskip \\
 % $3680\pm390$
\multirow{2}{*}{10NE}& \multirow{2}{*}{00:47:33.168} & \multirow{2}{*}{-25:17:16.89} & H30$\alpha$ &  $291\pm12$ & $1.94\pm0.28$ & $188\pm29$ & $387\pm87$ & 
%\multirow{2}{*}{
%\bigg\{}
\multirow{2}{*}{$1.80\pm0.57$, $2.91\pm0.83$}
\smallskip \\
 & & &   H26$\alpha$&  $257.0\pm4.7$ & $4.82\pm0.32$ & $161\pm11$ & $827\pm83$ 
 \medskip \\
 \multirow{2}{*}{11}& \multirow{2}{*}{00:47:33.172} & \multirow{2}{*}{-25:17:17.42} & H30$\alpha$ & $135.0\pm7.2$
 &  $2.31\pm0.39$
 &  $144\pm17$
 & $352\pm78$ 
 & \multirow{2}{*}{$1.1\pm0.59$} \smallskip \\
 & & &  H26$\alpha$
 & $135.1\pm4.5$ 
 & $4.00\pm0.36$ 
 & $133\pm11$ 
 & $560\pm73$ 
 & \medskip \\
 \multirow{2}{*}{13}& \multirow{2}{*}{00:47:33.193} &\multirow{2}{*}{-25:17:16.77} &  H30$\alpha$ &  $243.6\pm9.2$ & $2.80\pm0.73$ & $60\pm22$ & $179\pm85$ & \multirow{2}{*}{$1.0\pm1.3$}  \smallskip \\
 & & & H26$\alpha$& $235.2\pm5.8$ & $3.90\pm0.71$ & $67\pm14$ & $276\pm80$\\
\enddata
\tablenotetext{a}{Uncertainties of the integrated line intensities are estimated taking into account the uncertainties introduced by the baseline fit.}
\tablenotetext{b}{\mbox{RLSI}s derived from the velocity-integrated intensities of the Gaussian fits. In the case of source 10NE, the second shown value is that obtained from integrating the RL intensity just in the spectral range between 260 and \mbox{290 km s$^{-1}$} (one channel). We stress that, in the case of the $\alpha_\mathrm{L}$ derived for this spectral channel, we have not considered the uncertainty introduced by the baseline fit.}
\end{deluxetable*}

\subsection{Recombination Lines Spectral Indices}

Continuum sources do not spatially match the positions where the velocity-integrated submm RL emission peaks (see also \citealt{Mohan2005}). This is clearly illustrated in \mbox{Fig.~\ref{figure_with_maps}} where the continuum map at \mbox{231 GHz} has a significant/dominant contribution of dust emission, which does not trace the regions responsible for the free-free and RL emission. Dust emission dominating the free-free emission would also explain the high continuum spectral indices measured, $\alpha_\mathrm{c}\sim 2.7$--$4$, which are clearly above the value of 2 expected for an optically thick \mbox{H {\sc ii}} region. In addition, there are not reported measurements of the free-free continuum emission at multiple frequencies with the VLA toward these regions (note that sources 10, 11, and 13 were detected at 36 GHz by \citealt{Leroy2018} but the measurements at lower frequencies by \citealt{Ulvestad1997} are dominated by non-LTE emission). Therefore, the continuum maps cannot be used to infer the bremsstrahlung SED at the frequencies of the H30$\alpha$ and H26$\alpha$ RLs.

Fortunately, RL spectral indices (hereafter \mbox{RLSI}s),  $S_\mathrm{L} \Delta v \propto \nu^{\alpha_\mathrm{L}}$, are excellent tools to constrain the properties of the underlying \mbox{H {\sc ii}} regions. A clear example of the potential of using \mbox{RLSI}s is source 10, which shows non-thermal radio continuum emission and is thought to be the AGN of the galaxy \citep{Ulvestad1997}. For the case of optically thin emission, the expected \mbox{RLSI} is $\alpha_\mathrm{L}=1$ \citep{Altenhoff1960,Mezger1967}.  Larger values would indicate (partially) optically thick thermal emission in the lines. Special cases are the \mbox{H {\sc ii}} regions with electron density gradients. For the case of an isothermal ionized wind with an electron density, $N_\mathrm{e}$, that decreases with a power-law function of the radius, i.e. $N_\mathrm{e}\propto r^{-b_\mathrm{d}}$, the continuum flux follows a power-law dependence with the frequency as $S\propto \nu^{\alpha_\mathrm{c}}$, where the continuum spectral index $\alpha_\mathrm{c}$ is a function of $b_\mathrm{d}$ as illustrated in \mbox{Fig.~\ref{figure_of_alpha_vs_bd}} (see also \citealt{Panagia1975}). The \mbox{RLSI} follows the dependence with $b_\mathrm{d}$ shown also in \mbox{Fig.~\ref{figure_of_alpha_vs_bd}} in accordance with that calculated using the MORELI code, a 3D radiative transfer model for RLs \citep{BaezRubio2013}.  A measured value for the \mbox{RLSI}, $\alpha_\mathrm{L}$, larger than $\approx2.0\pm0.1$ would indicate RL maser emission once taken into account the uncertainty introduced by the electron temperature, whose value might reach \mbox{12000 K} in the extreme case of \mbox{H {\sc ii}} regions located at large galactocentric radii  \citep{Afflerbach1996,Quireza2006}.

\begin{figure}
\centering
\includegraphics[width=0.44\textwidth]{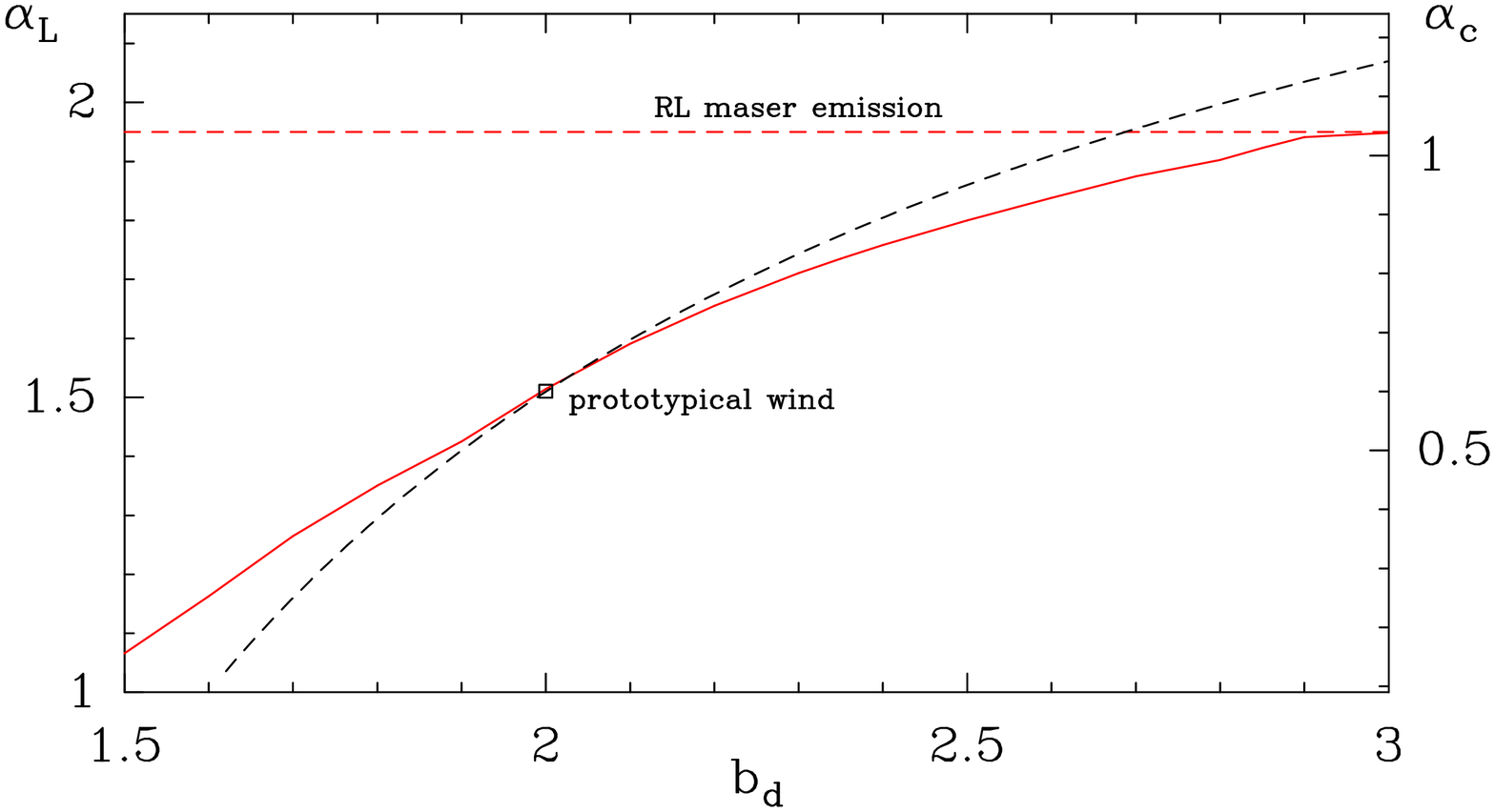}
\caption{Continuum and RL spectral indices (black dashed and red solid lines respectively) versus the index $b_\mathrm{d}$ of the power-law electron density gradient of an isothermal spherical ionized wind. For the calculations we have used a mass-loss rate of \mbox{$5\cdot10^{-5}$ M$_\odot$ yr$^{-1}$}, similar to that derived for MWC349A, an electron temperature of \mbox{6000 K} and an expansion velocity of \mbox{30 km s$^{-1}$ .} We have used the MORELI code \citep{BaezRubio2013} to take into account both the continuum and RL opacity effects. The open square shows the values found for a prototypical ionized wind with constant expansion velocity. The dashed horizontal line shows the boundary between thermal and RL maser emission. Note that this corresponds to $b_\mathrm{d}\approx3$.}
\label{figure_of_alpha_vs_bd}%
\end{figure}

\section{Constraining the ionized structure of the detected sources}
\label{results_section}

We have used \mbox{Fig.~\ref{figure_of_alpha_vs_bd}} and the measured \mbox{RLSI}s (see \mbox{Table 1)} to constrain the structure of the \mbox{H {\sc ii}} regions observed in \mbox{NGC 253.} The \mbox{RLSI}s from the two RLs have been derived from the H26$\alpha$ cube (with an original resolution of $0^{\prime\prime}23$) spatially smoothed to match the resolution of the H30$\alpha$ cube, i.e. $0.^{\prime\prime}303\times0.^{\prime\prime}248$ with \mbox{P.A.=-75.5$\degr$.} This allows us to analyze the ionized gas with a spatial resolution of about \mbox{5 pc.} The \mbox{RLSI}s were measured integrating the RL emission over the regions defined by their FWHM sizes derived from bidimensional Gaussian fits to the velocity-integrated RL intensities.  

The derived \mbox{RLSI}s toward our sources show values of $\approx$1.1 (sources 11 and 13), 2.15 (source 10), and 1.8--2.9 (source 10NE for two different ranges of spectral channels). As seen in \mbox{Fig.~\ref{figure_of_alpha_vs_bd},} the RL radiation is emitted by \mbox{H {\sc ii}} regions with very different electron density structures. We note that optically thin emission (source 13) with $\alpha_\mathrm{c}\approx-0.1$ and $\alpha_\mathrm{L}\approx1.0$ is expected to arise from the more evolved extended \mbox{H {\sc ii}} with moderate electron densities, while the partially optically thick emission with $\alpha_\mathrm{c}\sim0.6$ is expected to arise in younger and denser ultra and hyper-compact \mbox{H {\sc ii}} regions \citep{Jaffe1999}.

\subsection{Recombination line masers toward sources 10 and 10NE}

In contrast to sources 11 and 13, source 10  shows a \mbox{RLSI} of 2.15 (see \mbox{Table 1),} which is larger than the expected value for the thermal emission from any ionized thermal structure. The most plausible explanation is maser amplification of, at least, the H26$\alpha$ RL emission. This would provide a straightforward explanation of the large \mbox{RLSI} found between the H30$\alpha$ and H26$\alpha$ and their different line widths (which is mainly due to the lack of H30$\alpha$ RL emission in the spectral channels between 294 and \mbox{310 km s$^{-1}$}. However, given the uncertainties in the H30$\alpha$ detection, we cannot rule out that emission is under LTE conditions for an electron density profile with $b_\mathrm{d}>2$. 

On the other hand, a detailed analysis of the line profiles of 10NE between 260 and \mbox{290 km s$^{-1}$} reveals line intensities (integrated over the source) of $145\pm18$ and \mbox{$43\pm14$ mJy km s$^{-1}$} for the H26$\alpha$ and H30$\alpha$ RLs, respectively, at the 1$\sigma$ significance.
%%if one neglects uncertainties introduced by the baseline, given the narrow width of the spectral channel. 
Thus, using error propagation, we measure a RLSI of \mbox{$\alpha_\mathrm{L}=2.91\pm0.83$} (see \mbox{Table 1)}, which is above the upper limit for LTE emission, suggesting maser emission. For example, in the representative case of an ionized wind with an electron density of $b_\mathrm{d}=2$ (see, e.g. \citealt{Jaffe1999}), the expected RLSI for LTE emission would be about 1.5 (see \mbox{Fig.  \ref{figure_of_alpha_vs_bd})} and, therefore, our measured RLSI would indicate maser emission with a confidence level of about a 90\% (1.7 sigma-level). Only for the case of $b_\mathrm{d}=3$, the RLSI would be barely consistent with LTE emission at a 1 sigma-level.
%% if one assumes a steep electron density gradient of $b_\mathrm{d}=3$.
%% , giving as a result a probability of a 90\% that our claim results valid. 
However, we stress that this is an extreme case because, to the best  of our knowledge, the steepest electron density gradient known so far was found toward the evolved Wolf-Rayet galaxy of \mbox{NGC 4214 S,} which is very unlikely to be the case of \mbox{NGC 253,} showing a value of $b_\mathrm{d}\approx2.8$ \mbox{(see \citealt{Beck2000})}. Even in this unlikely case, the measured RLSI would indicate maser emission with a confidence level about 75\% (1.2 sigma-level).
Maser emission for the H26$\alpha$ RL is illustrated in  \mbox{Fig.~\ref{figure_with_maps}} by weak radio continuum and H30$\alpha$ emission combined with bright H26$\alpha$ emission seen in this source. We stress that RL opacities larger than those obtained in our model (e.g. due to dense clumps with optically thicker RL emission) would counteract the maser effect decreasing the measured \mbox{RLSI}s (see \citealt{Strelnitski1996b,Peters2012}). 

\subsection{Comparison with galactic RL masers.}

An argument in favor of maser emission in \mbox{NGC 253} is that one of the few known sources emitting hydrogen RL maser emission in our galaxy, Monoceros \mbox{R2-IRS2,} shows  $\alpha_\mathrm{L}\approx 1.81$ \citep{JimenezSerra2013}, a lower \mbox{RLSI} than that measured toward the two mentioned extragalactic sources. In addition, \mbox{MWC349A} shows a \mbox{RLSI} of 3.3 \citep{Thum1994a}, which is consistent (within the uncertainties) with the value derived toward source 10NE. Therefore, the measured $\alpha_\mathrm{L}$ toward source 10NE in \mbox{NGC 253} suggests that masers might arise from ionized winds similar to those in Monoceros R2-IRS2 and \mbox{MWC349A.} In fact, by scaling the fluxes (\mbox{$\sim30$ Jy for the H30$\alpha$ RL;} \citealt{Thum1994a}) from the actual distance of \mbox{MWC349A} (\mbox{1620 pc;} \citealt{Gaia2016,Gaia2018,Luri2018}) to the distance of \mbox{NGC 253} \mbox{(3.5 Mpc),} one predicts a H26$\alpha$ peak intensity of just \mbox{6.5 $\mathrm{\mu Jy}$} in \mbox{30 km s$^{-1}$}. Thus, since the measured peak intensity is \mbox{4.8 mJy} in 10NE, $\sim700$ \mbox{MWC349A}-like ionized winds are required to explain the measured intensity of the H26$\alpha$ line in 10NE in a \mbox{30 km s$^{-1}$} velocity range.

\section{Revealing the evolution of young super star clusters}
\label{section_of_young_SSCs}

The measured line widths of \mbox{$\Delta v\sim$65--180 km s$^{-1}$} are significantly broader than those measured toward most of the galactic \mbox{H {\sc ii}} regions, \mbox{$\Delta v\sim 30$ km s$^{-1}$,} due to thermal/turbulent broadening \citep{Mezger1967}. The broad lines in \mbox{NGC 253} are likely due to the velocity dispersion from unresolved \mbox{H {\sc ii}} regions detected in our \mbox{5 pc} beam along the line of sight. Assuming an electron temperature of \mbox{6000 K} and LTE emission for the H30$\alpha$ RL (which implies a velocity-integrated line-to-continuum ratio of \mbox{$\sim$125 km s$^{-1}$} assuming LTE optically thin RL emission; see \citealt{Rodriguez2009}), the Lyman continuum production rate required to maintain the ionization would be \mbox{$\gtrsim3\cdot10^{51}$ s$^{-1}$} toward all our sources. This is equivalent to $\geq80$ stars of spectral type O5 \citep{Vacca1996}. This reveals that the detected RLs arise in SSCs, which are formed by $\sim10^5$ individual stars \citep{Leroy2018}.
Since RL masers are typically amplified by an order of magnitude with respect to the expected   LTE values, we do not expect that the detected RL maser emission is dominated by just a few \mbox{H {\sc ii}} regions contained in our clusters, i.e. maser emission must be common toward a relatively high fraction of the \mbox{H {\sc ii}} regions. Maser sources are scarce in young stellar objects, implying that the expected timescale when these objects show this type of emission must be rather short, with a derived upper limit of \mbox{$\sim10^5$ yr} (see \citealt{Wood1989,Jaffe1999,BaezRubiothesis}). Therefore, our results provide strong evidence that the star population within these SSC may be coeval at scales of \mbox{$\sim10^4$ yr.} 

Based on the ionized structure deduced from the \mbox{RLSI}s, we could try to infer the history of  massive  star formation in the inner part of the nucleus of \mbox{NGC 253} surrounding the AGN as traced by the non-thermal source. The massive star formation in the inner part of \mbox{NGC 253} seems to indicate an evolutionary trend,  where the outermost sources 11 and 13 seem to be the most evolved (since they are optically thin), and SSCs closer to the putative AGN, sources 10 and 10NE, are the most recent star formation events because they show optically thicker emission.

\section{Conclusions}
\label{conclusions_section}

  We used ALMA data to study the continuum and RL emission of the H30$\alpha$ and H26$\alpha$ toward \mbox{NGC 253.} Our results show that \mbox{RLSI}s provide a unique tool to study the structure of the \mbox{H {\sc ii}} regions whose free-free continuum emission is not well established. We used the \mbox{RLSIs} to provide evidence for the first known extragalactic object masering in the H26$\alpha$ line, reaching a confidence level of about a 90\% for the likely case of a SSC composed by prototypical ionized winds. We stress that in highly confused regions, like those found in the vicinity of AGNs, \mbox{RLSI}s provide information to understand star formation in the innermost regions of active galaxies. In particular, ALMA sensitivity will detect RL maser emission from nearby galaxies at the highest available resolution allowing us to locate and to study in detail the formation process and the evolution of the embedded stellar population in the SSCs. 
   
\acknowledgments

This paper makes use of the following ALMA data: ADS/JAO.ALMA\#2013.1.00191.S, ADS/JAO.ALMA\#\\ 2015.1.00274.S,  ADS/JAO.ALMA\#2015.1.01133.S. ALMA is a partnership of ESO (representing its member states), NSF (USA) and NINS (Japan), together with NRC (Canada), MOST and ASIAA (Taiwan), and KASI (Republic of Korea), in cooperation with the Republic of Chile. The Joint ALMA Observatory is operated by ESO, AUI/NRAO and NAOJ. We acknowledge partial support by the MINECO and FEDER funding under grants ESP2015-65597-C4-1-R and ESP2017-86582-C4-1-R and by the Comunidad de Madrid grant no. S2013/ICE-2822 SpaceTec-CM. F.R.-V. also acknowledges support by the MINECO through grant BES-2016-078808. Finally, we are also grateful to the anonymous referee for
valuable comments.

\vspace{5mm}
\facilities{ALMA}

\software{CASA \citep{McMullin2007},  
          MADCUBA (\url{http://cab.inta-csic.es/madcuba/Portada.html}),
          MORELI \citep{BaezRubio2013}, 
          STATCONT \citep{SanchezMonge2018}
          }


\begin{thebibliography}{}
\expandafter\ifx\csname natexlab\endcsname\relax\def\natexlab#1{#1}\fi

\bibitem[{{Afflerbach} {et~al.}(1996){Afflerbach}, {Churchwell}, {Acord},
  {Hofner}, {Kurtz}, \& {Depree}}]{Afflerbach1996}
{Afflerbach}, A., {Churchwell}, E., {Acord}, J.~M., {et~al.} 1996, The
  Astrophysical Journal Supplement Series, 106, 423


\bibitem[{{Aleman} {et~al.}(2018){Aleman}, {Exter}, {Ueta}, {Walton},
  {Tielens}, {Zijlstra}, {Montez}, {Abraham}, {Otsuka}, {Beaklini}, {van Hoof},
  {Villaver}, {Leal-Ferreira}, {Mendoza}, \& {L{\'e}pine}}]{Aleman2018}
{Aleman}, I., {Exter}, K., {Ueta}, T., {et~al.} 2018, \mnras, 477, 4499

\bibitem[{{Altenhoff} {et~al.}(1960){Altenhoff}, {Mezger}, {Wendker}, \&
  {Westerhout}}]{Altenhoff1960}
{Altenhoff}, W., {Mezger}, P.~G., {Wendker}, H., \& {Westerhout}, G. 1960,
  Ver\"off. Univ.-Sternwarte Bonn, 59, 48

\bibitem[{B\'aez-Rubio(2015)}]{BaezRubiothesis}
B\'aez-Rubio, A. 2015, Ph.D. Thesis (UCM)

\bibitem[{{B{\'a}ez-Rubio} {et~al.}(2013){B{\'a}ez-Rubio},
  {Mart{\'{\i}}n-Pintado}, {Thum}, \& {Planesas}}]{BaezRubio2013}
{B{\'a}ez-Rubio}, A., {Mart{\'{\i}}n-Pintado}, J., {Thum}, C., \& {Planesas},
  P. 2013, \aap, 553, A45

\bibitem[{{B{\'a}ez-Rubio} {et~al.}(2014){B{\'a}ez-Rubio},
  {Mart{\'\i}n-Pintado}, {Thum}, {Planesas}, \&
  {Torres-Redondo}}]{BaezRubio2014}
{B{\'a}ez-Rubio}, A., {Mart{\'\i}n-Pintado}, J., {Thum}, C., {Planesas}, P., \&
  {Torres-Redondo}, J. 2014, \aap, 571, L4

\bibitem[{{Beck} {et~al.}(2000){Beck}, {Turner}, \& {Kovo}}]{Beck2000}
{Beck}, S.~C., {Turner}, J.~L., \& {Kovo}, O. 2000, \aj, 120, 244

\bibitem[{{Bendo} {et~al.}(2015){Bendo}, {Beswick}, {D'Cruze}, {Dickinson},
  {Fuller}, \& {Muxlow}}]{Bendo2015}
{Bendo}, G.~J., {Beswick}, R.~J., {D'Cruze}, M.~J., {et~al.} 2015, \mnras, 450,
  L80

\bibitem[{{Cox} {et~al.}(1995){Cox}, {Mart\'in-Pintado}, {Bachiller}, {Bronfman},
  {Cernicharo}, {Nyman}, \& {Roelfsema}}]{Cox1995}
{Cox}, P., {Mart\'in-Pintado}, J., {Bachiller}, R., {et~al.} 1995, \aap, 295, L39

\bibitem[{{Danchi} {et~al.}(2001){Danchi}, {Tuthill}, \&
  {Monnier}}]{Danchi2001}
{Danchi}, W.~C., {Tuthill}, P.~G., \& {Monnier}, J.~D. 2001, \apj, 562, 440

\bibitem[{{Gaia Collaboration} {et~al.}(2016){Gaia Collaboration}, {Prusti},
  {de Bruijne}, {Brown}, {Vallenari}, {Babusiaux}, {Bailer-Jones}, {Bastian},
  {Biermann}, {Evans}, \& et~al.}]{Gaia2016}
{Gaia Collaboration}, {Prusti}, T., {de Bruijne}, J.~H.~J., {et~al.} 2016,
  \aap, 595, A1

\bibitem[{{Gaia Collaboration} {et~al.}(2018){Gaia Collaboration}, {Brown},
  {Vallenari}, {Prusti}, {de Bruijne}, {Babusiaux}, {Bailer-Jones}, {Biermann},
  {Evans}, {Eyer}, \& et~al.}]{Gaia2018}
{Gaia Collaboration}, {Brown}, A.~G.~A., {Vallenari}, A., {et~al.} 2018, \aap,
  616, A1

\bibitem[{{Jaffe} \& {Mart{\'{\i}}n-Pintado}(1999)}]{Jaffe1999}
{Jaffe}, D.~T., \& {Mart{\'{\i}}n-Pintado}, J. 1999, \apj, 520, 162

\bibitem[{{Jim{\'e}nez-Serra} {et~al.}(2013){Jim{\'e}nez-Serra},
  {B{\'a}ez-Rubio}, {Rivilla}, {Mart{\'{\i}}n-Pintado}, {Zhang}, {Dierickx}, \&
  {Patel}}]{JimenezSerra2013}
{Jim{\'e}nez-Serra}, I., {B{\'a}ez-Rubio}, A., {Rivilla}, V.~M., {et~al.} 2013,
  \apjl, 764, L4

\bibitem[{{Jim{\'e}nez-Serra} {et~al.}(2011){Jim{\'e}nez-Serra},
  {Mart{\'{\i}}n-Pintado}, {B{\'a}ez-Rubio}, {Patel}, \&
  {Thum}}]{JimenezSerra2011}
{Jim{\'e}nez-Serra}, I., {Mart{\'{\i}}n-Pintado}, J., {B{\'a}ez-Rubio}, A.,
  {Patel}, N., \& {Thum}, C. 2011, \apjl, 732, L27

\bibitem[{{Leroy} {et~al.}(2018){Leroy}, {Bolatto}, {Ostriker}, {Walter},
  {Gorski}, {Ginsburg}, {Krieger}, {Meier}, {Mills}, {Ott}, {Rosolowsky},
  {Thompson}, {Veilleux}, \& {Zschaechner}}]{Leroy2018}
{Leroy}, A.~K., {Bolatto}, A.~D., {Ostriker}, E.~C., {et~al.} 2018, ArXiv
  e-prints, arXiv:1804.02083

\bibitem[{{Luri} {et~al.}(2018){Luri}, {Brown}, {Sarro}, {Arenou},
  {Bailer-Jones}, {Castro-Ginard}, {de Bruijne}, {Prusti}, {Babusiaux}, \&
  {Delgado}}]{Luri2018}
{Luri}, X., {Brown}, A.~G.~A., {Sarro}, L.~M., {et~al.} 2018, \aap, 616, A9

\bibitem[{{Mart{\'{\i}}n-Dom{\'e}nech}
  {et~al.}(2017){Mart{\'{\i}}n-Dom{\'e}nech}, {Rivilla}, {Jim{\'e}nez-Serra},
  {Qu{\'e}nard}, {Testi}, \& {Mart{\'{\i}}n-Pintado}}]{MartinDomenech2017}
{Mart{\'{\i}}n-Dom{\'e}nech}, R., {Rivilla}, V.~M., {Jim{\'e}nez-Serra}, I.,
  {et~al.} 2017, \mnras, 469, 2230

\bibitem[{{Mart\'in-Pintado} {et~al.}(1989){Mart\'in-Pintado}, {Bachiller},
  {Thum}, \& {Walmsley}}]{MartinPintado1989}
{Mart\'in-Pintado}, J., {Bachiller}, R., {Thum}, C., \& {Walmsley}, M. 1989,
  \aap, 215, L13

\bibitem[{{Mart{\'{\i}}n-Pintado} {et~al.}(2011){Mart{\'{\i}}n-Pintado},
  {Thum}, {Planesas}, \& {B{\'a}ez-Rubio}}]{MartinPintado2011}
{Mart{\'{\i}}n-Pintado}, J., {Thum}, C., {Planesas}, P., \& {B{\'a}ez-Rubio},
  A. 2011, \aap, 530, L15

\bibitem[{{McMullin} {et~al.}(2007){McMullin}, {Waters}, {Schiebel}, {Young},
  \& {Golap}}]{McMullin2007}
{McMullin}, J.~P., {Waters}, B., {Schiebel}, D., {Young}, W., \& {Golap}, K.
  2007, in Astronomical Society of the Pacific Conference Series, Vol. 376,
  Astronomical Data Analysis Software and Systems XVI, ed. R.~A. {Shaw},
  F.~{Hill}, \& D.~J. {Bell}, 127

\bibitem[{{Mezger} \& {Hoglund}(1967)}]{Mezger1967}
{Mezger}, P.~G., \& {Hoglund}, B. 1967, \apj, 147, 490

\bibitem[{{Mohan} {et~al.}(2005){Mohan}, {Goss}, \&
  {Anantharamaiah}}]{Mohan2005}
{Mohan}, N.~R., {Goss}, W.~M., \& {Anantharamaiah}, K.~R. 2005, \aap, 432, 1

\bibitem[{{Mouhcine} {et~al.}(2005){Mouhcine}, {Ferguson}, {Rich}, {Brown}, \&
  {Smith}}]{Mouhcine2005}
{Mouhcine}, M., {Ferguson}, H.~C., {Rich}, R.~M., {Brown}, T.~M., \& {Smith},
  T.~E. 2005, \apj, 633, 810

\bibitem[{{Olnon}(1975)}]{Olnon1975}
{Olnon}, F.~M. 1975, \aap, 39, 217

\bibitem[{{Panagia} \& {Felli}(1975)}]{Panagia1975}
{Panagia}, N., \& {Felli}, M. 1975, \aap, 39, 1

\bibitem[{{Peters} {et~al.}(2012){Peters}, {Longmore}, \&
  {Dullemond}}]{Peters2012}
{Peters}, T., {Longmore}, S.~N., \& {Dullemond}, C.~P. 2012, \mnras, 425, 2352

\bibitem[{{Planesas} {et~al.}(1992){Planesas}, {Mart{\'{\i}}n-Pintado}, \&
  {Serabyn}}]{Planesas1992}
{Planesas}, P., {Mart{\'{\i}}n-Pintado}, J., \& {Serabyn}, E. 1992, \apjl, 386,
  L23

\bibitem[{{Puche} {et~al.}(1991){Puche}, {Carignan}, \& {van
  Gorkom}}]{Puche1991}
{Puche}, D., {Carignan}, C., \& {van Gorkom}, J.~H. 1991, \aj, 101, 456

\bibitem[{{Rekola} {et~al.}(2005){Rekola}, {Richer}, {McCall}, {Valtonen},
  {Kotilainen}, \& {Flynn}}]{Rekola2005}
{Rekola}, R., {Richer}, M.~G., {McCall}, M.~L., {et~al.} 2005, \mnras, 361, 330

\bibitem[{{Quireza} {et~al.}(2006){Quireza}, {Rood}, {Bania}, {Balser}, \&
  {Maciel}}]{Quireza2006}
{Quireza}, C., {Rood}, R.~T., {Bania}, T.~M., {Balser}, D.~S., \& {Maciel},
  W.~J. 2006, \apj, 653, 1226

\bibitem[{{Rodr{\'{\i}}guez} {et~al.}(2009){Rodr{\'{\i}}guez}, {Zapata}, \&
  {Ho}}]{Rodriguez2009}
{Rodr{\'{\i}}guez}, L.~F., {Zapata}, L.~A., \& {Ho}, P.~T.~P. 2009, \apj, 692,
  162

\bibitem[{{S{\'a}nchez Contreras} {et~al.}(2017){S{\'a}nchez Contreras},
  {B{\'a}ez-Rubio}, {Alcolea}, {Bujarrabal}, \&
  {Mart{\'{\i}}n-Pintado}}]{SanchezContreras2017}
{S{\'a}nchez Contreras}, C., {B{\'a}ez-Rubio}, A., {Alcolea}, J., {Bujarrabal},
  V., \& {Mart{\'{\i}}n-Pintado}, J. 2017, \aap, 603, A67

\bibitem[{{S{\'a}nchez-Monge} {et~al.}(2018){S{\'a}nchez-Monge}, {Schilke},
  {Ginsburg}, {Cesaroni}, \& {Schmiedeke}}]{SanchezMonge2018}
{S{\'a}nchez-Monge}, {\'A}., {Schilke}, P., {Ginsburg}, A., {Cesaroni}, R., \&
  {Schmiedeke}, A. 2018, \aap, 609, A101

\bibitem[{{Smith} {et~al.}(1997){Smith}, {Strelnitski}, {Miles}, {Kelly}, \&
  {Lacy}}]{Smith1997}
{Smith}, H.~A., {Strelnitski}, V., {Miles}, J.~W., {Kelly}, D.~M., \& {Lacy},
  J.~H. 1997, \aj, 114, 2658

\bibitem[{{Strelnitski} {et~al.}(1996{\natexlab{a}}){Strelnitski}, {Haas},
  {Smith}, {Erickson}, {Colgan}, \& {Hollenbach}}]{Strelnitski1996c}
{Strelnitski}, V., {Haas}, M.~R., {Smith}, H.~A., {et~al.} 1996{\natexlab{a}},
  Science, 272, 1459

\bibitem[{{Strelnitski} {et~al.}(1996{\natexlab{b}}){Strelnitski}, {Ponomarev},
  \& {Smith}}]{Strelnitski1996b}
{Strelnitski}, V.~S., {Ponomarev}, V.~O., \& {Smith}, H.~A. 1996{\natexlab{b}},
  \apj, 470, 1118

\bibitem[{{Strelnitski} {et~al.}(1996{\natexlab{c}}){Strelnitski}, {Smith}, \&
  {Ponomarev}}]{Strelnitski1996a}
{Strelnitski}, V.~S., {Smith}, H.~A., \& {Ponomarev}, V.~O. 1996{\natexlab{c}},
  \apj, 470, 1134

\bibitem[{{Tafoya} {et~al.}(2004){Tafoya}, {G{\'o}mez}, \&
  {Rodr{\'{\i}}guez}}]{Tafoya2004}
{Tafoya}, D., {G{\'o}mez}, Y., \& {Rodr{\'{\i}}guez}, L.~F. 2004, \apj, 610,
  827

\bibitem[{{Thum} {et~al.}(1998){Thum}, {Mart\'in-Pintado}, {Quirrenbach}, \&
  {Matthews}}]{Thum1998}
{Thum}, C., {Mart\'in-Pintado}, J., {Quirrenbach}, A., \& {Matthews}, H.~E. 1998,
  \aap, 333, L63

\bibitem[{{Thum} {et~al.}(1994){Thum}, {Matthews}, {Mart{\'{\i}}n-Pintado},
  {Serabyn}, {Planesas}, \& {Bachiller}}]{Thum1994a}
{Thum}, C., {Matthews}, H.~E., {Mart{\'{\i}}n-Pintado}, J., {et~al.} 1994,
  \aap, 283, 582

\bibitem[{{Ulvestad} \& {Antonucci}(1997)}]{Ulvestad1997}
{Ulvestad}, J.~S., \& {Antonucci}, R.~R.~J. 1997, \apj, 488, 621

\bibitem[{{Vacca} {et~al.}(1996){Vacca}, {Garmany}, \& {Shull}}]{Vacca1996}
{Vacca}, W.~D., {Garmany}, C.~D., \& {Shull}, J.~M. 1996, \apj, 460, 914

\bibitem[{{Wood} \& {Churchwell}(1989)}]{Wood1989}
{Wood}, D.~O.~S., \& {Churchwell}, E. 1989, \apj, 340, 265

\bibitem[{{Zhang} {et~al.}(2017){Zhang}, {Claus}, {Watson}, \&
  {Moran}}]{Zhang2017}
{Zhang}, Q., {Claus}, B., {Watson}, L., \& {Moran}, J. 2017, \apj, 837, 53

\end{thebibliography}
\end{document}